\documentclass[12pt]{article}

\usepackage{epsfig}

\newcommand{\beao}{\begin{eqnarray*}}
\newcommand{\eeao}{\end{eqnarray*}}
\newcommand{\bea}{\begin{eqnarray}}
\newcommand{\eea}{\end{eqnarray}}
\newcommand{\be}{\begin{equation}}
\newcommand{\ee}{\end{equation}}

\newcommand{\la}{\langle}
\newcommand{\ra}{\rangle}

\newcommand{\bchi}{\mbox{\boldmath$\chi$}}
\newcommand{\btau}{\mbox{\boldmath$\tau$}}
\newcommand{\bsigma}{\mbox{\boldmath$\sigma$}}
\newcommand{\balpha}{\mbox{\boldmath$\alpha$}}
\newcommand{\bnabla}{\mbox{\boldmath$\nabla$}}
\newcommand{\bcA}{\mbox{\boldmath${\cal A}$}}
\newcommand{\bcB}{\mbox{\boldmath${\cal B}$}}
\newcommand{\bsB}{\mbox{\boldmath${\sf B}$}}

\newfont{\bcalfont}{cmbsy10 scaled 1200}

\begin{document}

\renewcommand{\thefootnote}{\fnsymbol{footnote}}

\title{Resolution of Gauss' law in Yang-Mills theory \\
by Gauge Invariant Projection: \\[0.5cm]
Topology and Magnetic Monopoles\footnotemark[2]}

\author{H. Reinhardt\footnotemark[4] \\
Institut f\"ur Theoretische Physik, Universit\"at T\"ubingen, \\
Auf der Morgenstelle 14, D-72076 T\"ubingen, Germany} 
\date{}

\maketitle

\footnotetext[2]{Supported by Deutsche Forschungsgemeinschaft, DFG-Re 856/1-3}
\footnotetext[4]{Electronic address: reinhardt@uni-tuebingen.de}

\renewcommand{\thefootnote}{\arabic{footnote}}

\begin{abstract}

An efficient way of resolving Gauss' law in Yang-Mills theory is presented by
starting from the projected gauge invariant partition function and integrating
out one spatial field variable. In this way one obtains immediately the
description in terms of unconstrained gauge invariant variables which was previously
obtained by explicitly resolving Gauss' law in a modified axial gauge. In this
gauge, which is a variant of 't Hooft's Abelian gauges, magnetic monopoles occur.
It is shown how the Pontryagin index of the
gauge field is related to the magnetic charges. It turns out that the magnetic monopoles are sufficient to account for the non-trivial topological structure of the theory.       

\end{abstract}
\newpage

\section{Introduction}

Quantization of gauge theory can be accomplished in two basically different 
ways:
canonical quantization and path integral quantization. The equivalence of both
approaches has been established e.g.\ in refs.\ \cite{<R101>}, \cite{<R104>}.
The fundamental problem in dealing with quantum
Yang-Mills theory is the elimination of the unphysical gauge degrees of freedom.
This can be accomplished by either reformulating the theory in manifestly gauge
invariant variables \cite{<R105>} or by explicitly resolving the Gauss law
constraint \cite{<R106>}.

Recently, for Yang-Mills theory on a torus, a complete resolution of Gauss' law has been accomplished in a modified
axial gauge in both the canonical operator approach \cite{<R107>} and in the
Hamilton functional
integral approach \cite{<R108>}. These approaches end up with the description of
Yang-Mills
theory in terms of two transverse degrees of freedom (for each group generator)
and a reduced Abelian field, living in the Cartan subgroup and in $D - 1$
dimensions. This seems to be the minimum number of gauge invariant degrees of
freedom necessary to describe Yang-Mills theory, at least on a torus. In the
present paper I show that, by starting from the projected gauge invariant
partition function, a more efficient resolution of Gauss' law is achieved by
applying the Weyl integration formula to the integration over the gauge group.
The result agrees with that obtained by a resolution of Gauss' law in modified axial gauge
\cite{<R107>}, \cite{<R108>}. In this gauge, which is equivalent to the
so-called Polyakov gauge, and represents a variant of 't Hooft's
Abelian gauge \cite{<R109>}, magnetic monopoles arise. Lattice calculations 
\cite{<R110>} indicate that magnetic monopoles are possibly the
dominant infrared degrees of freedom, at least in the so-called maximal Abelian
gauge \cite{<R111>} and also in the Polyakov gauge \cite{XX}. 
There are, however, also critical remarks on Abelian projection 
\cite{<RR>}
which favour center dominance, where vortices rather than monopoles are the
relevant infrared degrees of freedom.

Recently much work has been devoted to the relation between magnetic
monopoles and instantons \cite{<R111A>}. In ref.\ \cite{<R112>} it was found 
that in the maximal
Abelian gauge a monopole trajectory goes around an instanton.\footnote
{In ref.\ \cite{<R113>} it was found that a monopole trajectory passes through
the center of each instanton. However, in this case a gauge was adopted for 
which the pertinent gauge functional diverges rather than becoming minimal.}
In the Polyakov gauge a monopole trajectory was found to pass 
through the center of the instanton \cite{<R115>}. The distribution of monopoles
in a dilute instanton gas was determined in ref.\ \cite{<R114>} in a different
Abelian gauge. Recent lattice calculations \cite{<R111A>}, \cite{<R116>}
indicate that there is an intricate relation between instantons and magnetic
monopoles.

The essential features of the instantons are their topological properties, which
are measured by the Pontryagin index. Since monopoles are the sources of long
range fields they should influence the topology of the gauge fields. In this
paper I derive an exact relation between the Pontryagin index of the gauge field
and the magnetic charges of the monopoles contained in the gauge field.

The organization of the paper is as follows: In sect.\ 2 I start from
Yang-Mills theory in the Weyl gauge and define the gauge
invariant partition function. Gauge invariance is achieved here by integrating 
over all gauge equivalent initial states with the Haar measure of the gauge 
group. I consider then the path integral representation of the gauge invariant 
partition function and convert it to the standard Yang-Mills functional integral by a 
time-dependent gauge transformation, where the latter needs to have zero
winding number. This is shown in sect.\ 3, where the topology of the gauge field
is considered. In sect.\ 4, I perform a Cartan decomposition of the gauge group.
By applying the Weyl integration formula, the integration over the coset $SU
(N)/U (1)^{N-1}$ can be explicitly performed, leaving from the gauge invariant
projection a residual integration over the Cartan subgroup. Integrating out one
spatial component of the gauge field I obtain the desired representation of the
Yang-Mills partition function in gauge fixed, unconstrained variables, which was
previously derived by resolving Gauss' law in a modified axial gauge
\cite{<R107>}, \cite{<R108>}. In sect.\ 5, I discuss the emergence of magnetic
monopoles as a consequence of the Cartan decomposition of the gauge group and
performing a coset gauge transformation. Finally, in sect.\ 6 the relation
between the Pontryagin index and the magnetic charges of the monopoles is
derived. A short summary and some concluding remarks are given in sect.\ 7.  
A few
generic examples of monopole type of fields induced by singular gauge
transformations are presented in the appendix.

\section{The Gauge Invariant Partition Function}

We consider Yang-Mills theory with the gauge group $G = SU (N)$. The quantum
theory is defined in the Weyl gauge $A_0 = 0$ by the Hamiltonian 
\begin{eqnarray}
\label{<1>}
H = \int d^3 x \left( \frac{g^2}{2} E_i^a (x) E_i^a (x) + \frac{1}{2g^2}
 B_i^a (x) B_i^a (x) \right)\,. 
\end{eqnarray}

Here, the electric field $E_k^a (x) = \frac{1}{i} \frac{\delta}{\delta
A_k^a (x)}$ is the canonical momentum conjugate to the field coordinate
$A_k^a (x)$  and $B_k^a = \epsilon_{k i j} (\partial_i A_j^a + \frac{1}{2}
f^{abc} A_i^b A_j^c)$ is the magnetic field. Furthermore, $g$ is the (bare) coupling
constant and $f^{abc}$ is the structure constant of the gauge group. 

Let $| C \ra$ denote an eigenstate of $A_i (x)$, i.e. 
\begin{eqnarray}
A_i (x) | C \ra = C_i (x) | C \ra \,, 
\end{eqnarray}

where $C_i (x)$ is some classical field function. The gauge invariant partition
function of Yang-Mills theory can then be defined as (see e.g.\ refs.\ \cite
{<R101>}, \cite{<R103>}, \cite{<R102>})
\begin{eqnarray}
\label{<3>}
Z = \int {\cal D} C_i (x) \la C | e^{-HT} P | C \ra \,, 
\end{eqnarray}

where the (functional) integration runs over all classical field functions $C_i
(x)$ and $P$ projects onto gauge invariant states:
\begin{eqnarray}
\label{<4>}
P | C \ra = \sum_n e^{-in \Theta} \int\limits_G {\cal D} \mu (\Omega_n) 
|C^{\Omega_n} \ra 
\end{eqnarray}

Here, $A_i^\Omega (x)$ denotes the gauge transform of the gauge 
field defined by\footnote{We are using antihermitian fields $A_\mu = A_\mu^a
T^a$ with generators $T^a$ satisfying $[T^a, T^b] = f^{abc} T^c$, 
$\mbox{Tr} (T^a T^b) = - \frac{1}{2} \delta^{ab}$. }
\begin{eqnarray}
A_i^\Omega = \Omega A_i \Omega^\dagger + \Omega \partial_i
\Omega^\dagger\,. 
\end{eqnarray} 

Furthermore, $\Theta$ is the vacuum angle \cite{<R117>} 
and $\mu (\Omega)$ denotes the
Haar measure of the gauge group. The integration runs over all time-independent
gauge transformations $\Omega_n$ with winding number $n$, which is defined by
\begin{eqnarray}
\label{<6>}
n [\Omega] = - \frac {1}{24 \pi^2} \int d^3 x \, \epsilon^{ijk} 
\mbox{Tr} L_i L_j L_\kappa\,, \quad
L_\kappa = \Omega \partial_\kappa \Omega^\dagger\,. 
\end{eqnarray} 

As usual we assume here that the gauge function $\Omega (\mathbf{x})$ approaches a
unique value $\Omega_\infty$ for $|\mathbf{x}| \rightarrow \infty$, so that
$R^3$ can be compactified to $S^3$ and $n [\Omega]$ is a topological invariant.
For later convenience we will choose 
\begin{eqnarray}
\label{<6A>}
\lim_{| \mathbf{x}| \to \infty}
\Omega_n (\mathbf{x}) = \Omega_\infty^n = (- 1)^n\,. 
\end{eqnarray}

Following the standard procedure \cite{<R118>} one derives the following functional
integral representation of the Yang-Mills transition amplitude:
\begin{eqnarray}
\label{<7>}
\la C | e^{-HT} | C' \ra = \int_{C'}^C {\cal D} A_i \, \exp (- S_{\mathrm{YM}} 
[A_0 = 0,
\mathbf{A}]) 
\end{eqnarray}
 
Here, the functional integration runs over all field configurations $A_i (x)$
satisfying the boundary conditions $A_i (x_0 = 0, \mathbf{x}) 
= C'_i (\mathbf{x})$,
$A_i (x_0 = T, \mathbf{x}) = C_i (\mathbf{x})$. Furthermore
\begin{eqnarray}
S_{\mathrm{YM}} [A_0, \mathbf{A}] = -\frac{1}{2g^2} \int d^4 x \mbox{Tr} F_{\mu \nu} (x) F_{\mu
\nu} (x) 
\end{eqnarray} 

is the standard Yang-Mills action with 
\begin{eqnarray}
F_{\mu \nu}^a = \partial_\mu A_\nu^a - \partial_\nu A_\mu^a + f^{abc} A_\mu^b
A_\nu^c 
\end{eqnarray} 

denoting the field strength. Inserting eq.\ (\ref{<7>}) into eq.\ (\ref{<3>}) we 
obtain for the partition function
\begin{eqnarray}
\label{<11>}
Z = \sum_n  e^{- in \Theta} \int\limits_G {\cal D}\mu (\Omega_n)
\int\limits_{C^\Omega}^C
{\cal D} A_i \hspace{0.3cm} e^{-S _{\mathrm{YM}} [A_0 = 0, \mathbf{A}]}\,, 
\end{eqnarray} 

where the functional integration is performed with boundary conditions
\begin{eqnarray}
\label{<11A>}
A_i (x_0 = 0, \mathbf{x}) = C_i^\Omega (\mathbf{x})\,, \quad 
A_i (x_0 = T, \mathbf{x}) 
= C_i (\mathbf{x})\,. 
\end{eqnarray} 

Note that equation (\ref{<11>}) is almost the standard Yang-Mills functional
integral representation except that, instead of an integral over the time
component of the gauge field $A_0$, we have the integration over the gauge
group with the Haar measure.\footnote{The importance of the Haar measure has
been emphasized in ref.\ \cite{<R102>}. Let me also mention that 
eq.\ (\ref{<11>}) was previously obtained in ref.\ \cite{RXX}.} The ultimate 
relation between the gauge
transformation and the time component of the gauge field can be easily
established by performing the time-dependent gauge transformation $A
\to A^U$ with
\begin{eqnarray}
\label{<12>}
U = \Omega_n^{\frac{x_o}{T} -1}\,, 
\end{eqnarray} 

which removes the gauge transformation $\Omega_n$ from the initial values of the
spatial gauge fields (\ref{<11A>}) but introduces at the same time a 
time-independent temporal gauge potential 
\begin{eqnarray}
\label{<13>}
A_0 = - \frac{1}{T} \ln \Omega_n\,. 
\end{eqnarray}

The partition function then becomes (for simplicity $A_\mu^U$ is replaced by
$A_\mu$)
\begin{eqnarray}
\label{<14>}
Z = \sum_n  e^{- in \Theta} \int\limits_G {\cal D} \mu (e^{-TA_0}) \int
{\cal D} A_i \, e^{-S_{\mathrm{YM}} [A_0, \mathbf{A}]}\,, 
\end{eqnarray}

where the functional integration now runs over temporally periodic spatial gauge
fields $A_i (T, \mathbf{x}) = A_i (0, \mathbf{x})$. This is almost the standard
Yang-Mills functional integral except for the presence of the Haar measure for
the temporal gauge field (\ref{<13>}) and the absence of the gauge fixing by the
Faddeev-Popov method. It has been shown \cite{<R104>}, however, that the 
representation (\ref{<14>}) is completely equivalent to the standard Yang-Mills 
functional integral, where the integration is performed with a flat measure over
all four components of the gauge field $A_\mu (x)$ but with the gauge fixed by 
the Faddeev-Popov method. In this case the Haar measure arises from the 
Faddeev-Popov determinant. In the equivalence proof given in ref.\ \cite{<R104>} it 
was tacitly assumed that the gauge transformation (\ref{<12>}) does not change the
topological properties of the gauge field. To be more precise, for 
$\Theta \neq 0$ the
equivalence of eq.\ (\ref{<14>}) with the standard Yang-Mills functional
integral requires that the winding number $n$ of the gauge function $\Omega$
coincides with the (negative) Pontryagin index
\begin{eqnarray}
\label{<R1>}
\nu [A] = - \frac{1}{16 \pi^2} \int d^4 x \mbox{Tr} F_{\mu \nu} 
\tilde{F}_{\mu \nu}\,,
\quad \tilde{F}_{\mu \nu} = \frac{1}{2} \epsilon_{\mu \nu \kappa \lambda} 
F_{\kappa \lambda} 
\end{eqnarray}

of the gauge rotated field $A_\mu^U$. This will be shown in the next section.

\section{Topology of the Gauge Field}

The gauge fields are topologically classified by the Pontryagin index
(\ref{<R1>}).
It is well known that the integrand of eq.\ (\ref{<R1>}) is a total derivative 
\begin{eqnarray}
\label{<R2>}
- \frac{1}{16 \pi^2} \, \mbox{Tr} F_{\mu \nu} \tilde{F}_{\mu \nu} 
= \partial_\mu X_\mu \,,
\end{eqnarray}

where
\begin{eqnarray}
\label{R3>}
X_\mu [A] = - \frac{1}{16 \pi^2} \epsilon_{\mu \nu \kappa \lambda} 
\mbox{Tr} 
\left\{ \frac{1}{2} A_\nu \partial_\kappa A_\lambda + \frac{1}{3} 
A_\nu A_\kappa A_\lambda \right\} 
\end{eqnarray}

is the topological current. For smooth gauge fields\footnote{We will assume
that the gauge fields $A_i (x)$ are smooth in the Weyl gauge.} we can apply
Gauss' theorem to obtain 
\begin{eqnarray}
\label{<R4>}
\nu [A] = \int_{\partial M} d \Sigma_\mu X_\mu [A]\,. 
\end{eqnarray}

Under a gauge transformation the topological current transforms as 
\begin{eqnarray}
\label{<R5>}
X_\mu [A^U] = X_\mu [A] + \frac{1}{16 \pi^2} \epsilon_{\mu \alpha \beta
\gamma} \mbox{Tr} \Big\{ 2 \partial_\beta (L_\alpha A_\gamma) 
+ 2 A_\alpha \ell_{\beta \gamma} + L_\alpha \ell_{\beta \gamma} - \frac{2}{3} L_\alpha
L_\beta L_\gamma \Big\} 
\end{eqnarray}

with the current
\begin{eqnarray}
\label{<R6>}
L_\alpha = U \partial_\alpha U^\dagger   
\end{eqnarray}

and
\begin{eqnarray}
\label{<R7>}
\ell_{\mu \nu} = U (\partial_\mu \partial_\nu - \partial_\nu \partial_\mu)
U^\dagger \,. 
\end{eqnarray}

The last quantity vanishes for non-singular gauge functions $U (x)$, which we
will assume. Furthermore, for smooth $U(x)$ the surface term in (\ref{<R5>}) will not
contribute and we obtain
\begin{eqnarray}
\label{<R8>}
\nu [A^U] = \nu [A] + \bar{n} [U] \,, 
\end{eqnarray}

where
\begin{eqnarray}
\label{<R9>}
\bar{n} [U] \equiv \nu [U \partial U^\dagger] = - \frac{1}{24 \pi^2} 
\int\limits_{\partial M} d^3 \Sigma_\mu \epsilon_{\mu \alpha \beta \gamma}
\mbox{Tr} (L_\alpha L_\beta L_\gamma) 
\end{eqnarray}

is the extension of the winding number (\ref{<6>}) to time-dependent gauge functions
$U (\mathbf{x}, t)$ (see below).

\begin{figure}
\begin{minipage}{7.5cm}
\begin{center}
\epsfig{file = 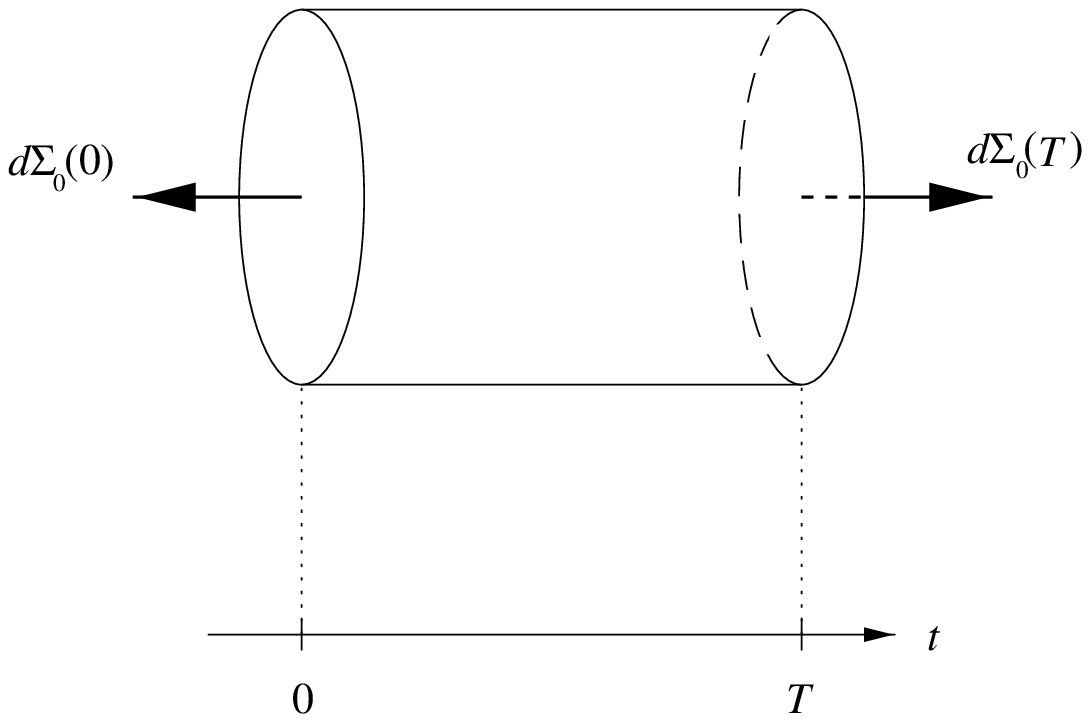, width = 6cm, height = 4cm} \\
(a) 
\end{center}
\end{minipage}
\hfill
\begin{minipage}{7.5cm}
\begin{center}
\epsfig{file = 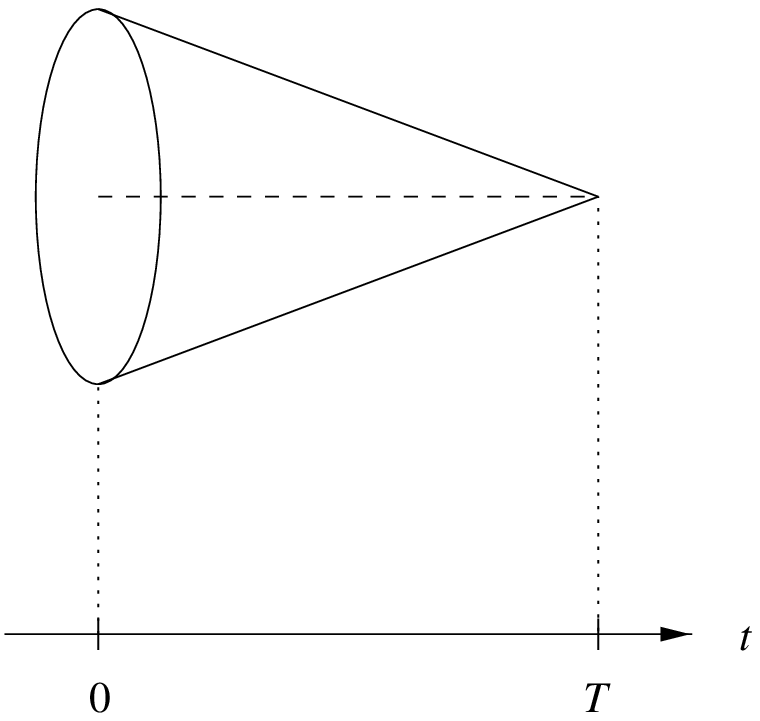, width = 4.5cm, height = 4cm} \\
(b) 
\end{center}
\end{minipage}
\caption{\label{fig1} Compactification of four-dimensional Euclidean space
from a cylinder (a)
to a cone (b) by contracting the face of the cylinder at $t = T$ to a single
point. }
\end{figure}

For sake of completeness let us first determine the Pontryagin index
(\ref{<R4>})
in the Weyl gauge $A_0 = 0$.
The integration in eq.\ (\ref{<R4>}) is over the surface
of four-dimensional Euclidean space, which, for a finite time interval $T$,
is given by a four-dimensional cylinder, $M=[0,1]\times B^3_\infty$, 
see fig.\ \ref{fig1}a. The faces of the cylinder are given by our ordinary
3-dimensional space, which we assume here to be a 3-dimensional ball 
$B^3_\infty$ of infinite radius.\footnote{In view of the boundary condition
(\ref{<6A>}) the 3-dimensional space $B^3_\infty$ can be topologically compactified
to a sphere $S^3_\infty$ of infinite radius. Let us emphasize that
compactification is understood here only in the topological sense but not in the
metrical sense.} 
The mantle
of the cylinder does not contribute to the Pontryagin index (\ref{<R4>}) in the
Weyl gauge provided the field configurations $A_i (x)$ drop off faster than $1/
|\mathbf{x}|$  at spatial infinity $|\mathbf{x}| \to \infty$.\footnote{This is 
the case in the absence of magnetic monopoles. We
will assume here that magnetic monopoles are absent in the Weyl gauge. Magnetic
monopoles will later on occur as gauge artifacts after performing singular gauge
transformations, i.e.\ by going from the smooth Weyl gauge to singular gauges.}
Furthermore, since 
\begin{eqnarray}
\label{<R10>}
d^3 \Sigma_0 (t = T) = - d^3 \Sigma_0 (t = 0) = d^3 x 
\end{eqnarray}

the contribution from the faces of the cylinder reads
\begin{eqnarray}
\label{<R11>}
\nu [A] = W [A (t = T)] - W [A (t = 0)]\,, 
\end{eqnarray}

where 
\begin{eqnarray}
\label{<R12>}
W [A] = \int d^3 x \, X_0 [A] 
\end{eqnarray}

is the charge of the topological current (\ref{<R5>}), which is nothing but the
Chern-Simons action. Therefore, for gauge fields which contribute to the
partition function, i.e.\ satisfy the boundary conditions (\ref{<11A>}), we 
obtain for the Pontryagin index 
\begin{eqnarray}
\label{<R13>}
\nu [A] = W [C] - W [C^\Omega]\,. 
\end{eqnarray}

From the transformation property of the topological current (\ref{<R5>}) we obtain
for smooth $\Omega (\mathbf{x})$ 
\begin{eqnarray}
\label{<R14>}
W [A^\Omega] = W [A] + n [\Omega] 
\end{eqnarray}

and from eq.\ (\ref{<R13>}), it follows that 
\begin{eqnarray}
\label{<R15>}
\nu [A] = - n [\Omega] \,,
\end{eqnarray}

where $n [\Omega]$ (\ref{<6>})
is the reduction of the winding number $\bar{n}$ (\ref{<R9>}) to 
time-independent $\Omega(\mathbf{x})$. 
It is this quantity $n [\Omega] \in \Pi_3 (S^3)$ which is
usually referred to as ``winding number''.

It remains to be shown how the Pontryagin index changes under the
time-dependent gauge transformation $U (x)$ (\ref{<12>}), see eq.\ (\ref{<R8>}). 
In this case the integration in the expression for the winding number
(\ref{<R9>}) is over the four-dimensional cylinder illustrated in 
fig.\ \ref{fig1}a. With
eq.\ (\ref{<R10>}) we can rewrite the winding number as 
\begin{eqnarray}
\label{<R16>}
\bar{n} [U] = \int d^3 x X_0 (t = T) - \int d^3 x X_0 (t = 0) 
+ \int\limits_{\mathcal{M}} d^3 \Sigma_i X_i 
\end{eqnarray}

where $\mathcal{M}$ denotes the mantle of the cylinder. Since $U (t = T) = 1$, the
integrand of the first term vanishes. Furthermore, since $U (t = 0) =
\Omega^{-1}$, the second term yields the contribution 
\begin{eqnarray}
\label{<R17>}
- n [\Omega^{-1}] = n [\Omega] \,.
\end{eqnarray}

Let us now calculate the contribution from the mantle of the cylinder (the last
term in eq.\ (\ref{<R16>})). 
Exploiting the properties of the anti-symmetric tensor $\epsilon_{\mu \alpha
\beta \gamma}$, the mantle contribution becomes
$(\epsilon_{0ijk} = \epsilon_{ijk})$
\begin{eqnarray}
\label{<R22>}
\bar{n}^{(\mathcal{M})} [U] = \frac{1}{16 \pi^2} \int\limits_{\mathcal{M}} 
d^3 \Sigma_i \epsilon_{ijk} \,\mbox{Tr}
(L_0 [L_j, L_k])\,. 
\end{eqnarray} 

Using $F_{ij} [L] = 0$, i.e.
\begin{eqnarray}
[L_i, L_j] = - (\partial_i L_j - \partial_j L_i) \,,
\label{neu3}
\end{eqnarray} 

we can rewrite the above expression as
\begin{eqnarray}
\label{<RX>}
\bar{n}^{(\mathcal{M})} [U] = - \frac{1}{8 \pi^2} \int\limits_{\mathcal{M}} 
d^3 \Sigma_i
\epsilon_{ijk} \mbox{Tr} (L_0 \partial_j L_k)\,. 
\end{eqnarray} 

To simplify the following calculations I will restrict myself to the gauge group
$G = SU(2)$.
Parameterizing the group element by ($T^a = - \frac{i}{2} \tau^a$, where 
$\tau^a$ are the Pauli matrices)
\begin{eqnarray}
\label{<R18>}
\Omega = \exp (i \bchi  \btau ) = \cos \chi + i \btau 
\hat{\bchi} \sin \chi\,, \quad
\chi = |\bchi|\,, \quad \hat{\bchi} = \bchi/\chi 
\end{eqnarray} 

we find 
\begin{eqnarray}
\label{<R19>}
U = \Omega^{\frac{t}{T} -1} = \cos \bar{\chi} - i \hat{\bchi}\btau \sin 
\bar{\chi} \,, \quad \bar{\chi} = \chi \left( 1 - \frac{t}{T} \right)
\end{eqnarray} 

and therefore
\begin{eqnarray}
\label{<R20>}
L_t = U \partial_t U^\dagger = - \frac{i}{T} \bchi  \btau\,. 
\end{eqnarray} 

For the spatial currents straightforward evaluation yields
\begin{eqnarray}
\label{R23>}
L_k &=& U \partial_k U^\dagger \nonumber\\ 
&=& i \hat{\bchi} \btau  \partial_k
\bar{\chi} + i \sin \bar{\chi} \cos \bar{\chi} \partial_k \hat{\bchi}
\cdot\btau 
+ i \sin^2 \bar{\chi} (\hat{\bchi} \times \partial_k \hat{\bchi}) 
\btau\,. 
\end{eqnarray} 

Hence we find
\begin{eqnarray}
\label{<R24>}
\epsilon_{ijk}\mbox{Tr} (L_0 \partial_j L_k) = 2\epsilon_{ijk} 
\frac{\chi}{T} \sin^2 \bar{\chi} (\partial_j \hat{\bchi} \times \partial_k 
\hat{\bchi}) \cdot \hat{\bchi}\,. 
\end{eqnarray} 

Inserting the last expression into eq.\ (\ref{<RX>}), we obtain for the mantle
contribution
\begin{eqnarray}
\bar{n}^{(\mathcal{M})} [U] = - \frac{1}{4 \pi^2} \int\limits_{\mathcal{M}} 
d^3 \Sigma_i \frac{\chi}{T} \sin^2
\bar{\chi} \epsilon_{ijk} (\partial_j \hat{\bchi} \times \partial_k 
\hat{\bchi}) \hat{\bchi}\,.
\end{eqnarray} 

The surface element of the mantle is given here by
\begin{eqnarray}
\label{<R21>}
d^3 \Sigma_i = d t d^2 \sigma_i \,, 
\end{eqnarray} 

where $d^2 \sigma_i$ denotes the surface element of three-dimensional space, 
i.e.\ the integration runs over $\mathcal{M} = S_\infty^2 \times [0, T]$ where
$S_\infty^2$ is the surface of our 3-space $B^3_\infty$.
In eq.\ (\ref{<R22>}) the
integrand has to be taken at spatial infinity, where, in view of the boundary
condition on $\Omega$ (\ref{<6A>}), 
\begin{eqnarray}
\label{<R25>}
\lim_{r \to \infty} \chi (r, \hat{\mathbf{x}}) = n \pi\,. 
\end{eqnarray}

The time integral can then be
readily performed, since $\hat{\bchi}$ is time-independent, yielding 
\begin{eqnarray}
\label{<R26>}
\frac{1}{T} \int\limits_0^T d t\, \chi \sin^2 \bar{\chi} = \frac{n \pi}{T}
\int\limits_0^T d t \sin^2 n \pi \left( 1 - \frac{t}{T} \right) = \int_0^{n \pi} dz 
\sin^2 {\it{z}} = \frac{n \pi}{2}\,. 
\end{eqnarray}

We therefore obtain for the contribution from the mantle of the cylinder
\begin{eqnarray}
\bar{n}^{(\mathcal{M})} [U] = -n m [\hat{\bchi}]\,,
\label{<45>}
\end{eqnarray}

where
\begin{eqnarray}
\label{<R27>}
m [\hat{\bchi}] = \frac{1}{8 \pi} \int\limits_{S_\infty^2} d \sigma_i \,
\epsilon_{ijk} (\partial_j \hat{\bchi} \times \partial_k  \hat{\bchi}) 
\hat{\bchi} 
\end{eqnarray}

defines the winding number $m[\hat{\bchi}] \in \Pi_2 (S^2)$ of the mapping 
$\hat{\bchi} (x)$ of the surface
$S_\infty^2$ of $R^3$ into the equator of the gauge group $\Omega
\left( \chi = \frac{\pi}{2} \right) = i \hat{\bchi}\btau$, which is
also $S^2$ (since $\hat{\bchi}^2 = 1$). 

By definition of homotopy classes mappings with the same winding number can be
smoothly deformed into each other. Therefore, we can smoothly map any gauge 
function $\Omega(\mathbf{x})$ onto the corresponding ``hedgehog'' map 
$\hat{\bchi} (\mathbf{x}) =\hat{\mathbf{x}}$ with the same winding number. 
Since $m [\hat{\mathbf{x}}] = 1$ we find from (\ref{<45>})
\begin{eqnarray}
\bar{n}^{(\mathcal{M})} [U] = - n [\Omega]\,, 
\end{eqnarray}

which together with eqs.\ (\ref{<R16>}), (\ref{<R17>}) implies
\begin{eqnarray}
\label{<R30>}
\bar{n} [U] = 0\,.  
\end{eqnarray}

This result can be easily understood from fig.\ \ref{fig1}. 
Since $U (t = T) = 1$, all
points of the face at $t = T$ can be identified. This deforms the cylinder to
the cone shown in fig.\ \ref{fig1}b. This cone can be thought of as a 
deformation of the
face at $t = 0$ (with opposite orientation), since both manifolds have the same
boundary, namely the boundary $S_\infty^2$ of 3-space. Furthermore, by our
choice of boundary conditions, $\Omega (\mathbf{x})$ traces out 
the same group space on both manifolds. However, the two manifolds have opposite orientations and
hence give opposite contributions to the winding number. 

With (\ref{<R30>}) we
finally obtain
\begin{eqnarray}
\label{<R15A>}
\nu [A^U] = \nu [A]\,,  
\end{eqnarray}

showing that the Pontryagin index does not change under the gauge transformation
$U (x)$ (\ref{<12>}). Therefore we can replace ($- n$) in eq.\ (\ref{<14>}) by 
$\nu [A]$ and
obtain the desired result.

\section{Cartan Decomposition of the Gauge Transformation}

It is convenient to perform a diagonalization of the map $\Omega_n (x)$ provided 
by the gauge function,
\begin{eqnarray}
\label{<15>}
\Omega (x) = V^\dagger \omega V \,, 
\end{eqnarray}

where $\omega$ is an element of the Cartan subgroup (maximal torus) $H = U
(1)^{N-1}$ and $V$ lives in the coset $G/H$. By gauge invariance of the
Yang-Mills
Hamiltonian, we have
\begin{eqnarray}
\label{<16>}
\la C | e^{- HT} | C^{V^\dagger \omega V} \ra = \la C^V | e^{- HT} |
(C^V)^\omega \ra\,. 
\end{eqnarray}

By shifting the integration invariable $C (x)^V \rightarrow C (x)$ it is seen
that the integrand in the partition function (\ref{<3>}), (\ref{<4>}) does not
depend on the coset $V$. 

The integration over the coset $G/H$ can then be explicitly performed by using the Weyl
integration formula \cite{<RY>}
\begin{eqnarray}
\label{<Z1>}
\int d \mu (\Omega_n) f (\Omega_n) = \frac{1}{| \mathcal{W} |} \int\limits_H d
\bar{\mu} (\omega_n) \int\limits_{G/H} d V_n f (V_n^\dagger \omega_n V_n)\,,
\end{eqnarray}

where $|\mathcal{W}|$ is the order of the Weyl group $(|\mathcal{W}| =
N!$ for $G = SU (N))$ and the reduced Haar measure $\bar{\mu} (\omega)$ is
defined by 
\begin{eqnarray}
\label{<Z2>}
d \bar{\mu} (\omega) = \prod_k d \lambda_k \, \sum_p \, 
\delta \left( \sum_i \lambda_i - 2 \pi p \right) \prod_{i < j}
\sin^2 \frac{\lambda_i - \lambda_j}{2}\,. 
\end{eqnarray}

Here $i \lambda_k$ denotes the eigenvalues of $\ln \omega_n$ ($\lambda_k$ 
real).\footnote{When the Weyl formula is extended to functional integrals over
two-dimensional compact manifolds, topological obstructions occur, as discussed
recently in ref.\ \cite{<R119>}. In higher dimensions these obstructions are
absent.} 
The partition function then becomes (cf.\ eqs.\ (\ref{<11>}) and
(\ref{<14>}))
\begin{eqnarray}
\label{<Z3>}
Z = \sum _n \quad e^{- in \Theta} \int\limits_H \mathcal{D} \bar{\mu} 
(e^{-Ta_0}) \int
\mathcal{D} A'_i (x) \exp (- S_{\mathrm{YM}} [a_0, \mathbf{A}'])\,, 
\end{eqnarray}

where
\begin{eqnarray}
\label{<Z4>}
a_0 = - \frac{1}{T} \ln \omega_n 
\end{eqnarray}

and an irrelevant constant, which arises from the integration over the coset
space $V_n$, has been dropped. The spatial
gauge fields $A'_i (x)$ in eq.\ (\ref{<Z3>}) are related to
the ones in eq.\ (\ref{<14>}) by the time-independent gauge transformation 
$V_n \in G/H$,
\begin{eqnarray}
\label{<Z5>}
A'_i = A^V_i = VA_i V^\dagger + V \partial_i V^\dagger\,, 
\end{eqnarray}

and hence the $A'_i (x)$ also fulfill temporally periodic boundary conditions
(we will omit the prime in the following). 

Since a specific Lorentz component of the gauge field $A_\mu (x)$, $\mu$
fixed, enters the Yang-Mills action $S_{\mathrm{YM}} [A_0, A_i]$ at most quadratically,
we can explicitly integrate out one spatial component of the gauge field, say $A_3 (x)$, 
and obtain 
\begin{eqnarray}
\label{<Z6>}
Z  = \sum_n e^{- in \Theta} \int\limits_H \mathcal{D} \mu (e^{- Ta_0}) 
\int \mathcal{D}  A_1 \mathcal{D} A_2  \mbox{Det}^{-1/2}
(- \hat{D}_{\bar{\mu}} \hat{D}_{\bar{\mu}})
\exp (- \tilde{S} (a_0, A_1, A_2))\,.   
\end{eqnarray}

Here
\begin{eqnarray}
\label{<Z7>}
\tilde{S} (A_0, A_1, A_2,) = \frac{1}{2g^2} \int d^4 x \, \partial_3 A_
{\bar{\mu}} \mathcal{P}_{\bar{\mu} {\bar{\nu}}} \partial_3  A_{\bar{\nu}}
+ \frac{1}{4 g^2} \int d^4 x
(F_{\bar{\mu} {\bar{\nu}}})^2 
\end{eqnarray}

with
\begin{eqnarray}
\label{<Z8>}
\mathcal{P}_{\bar{\mu} \bar{\nu}} = \delta_{\bar{\mu} \bar{\nu}} - 
\hat{D}_{\bar{\mu}} \,
\frac{1}{\hat{D}_{\bar{\lambda}} \hat{D}_{\bar{\lambda}}} \hat{
D}_{\bar{\nu}}\,, \quad \hat{D}_\mu^{ab} = \delta^{ab} \partial_\mu + f^{acb} 
A_\mu^c 
\end{eqnarray}

is the effective action of the remaining gauge degrees of freedom.
Note that the indices $\bar{\mu}, \bar{\nu}, \dots$ run only from 0 to 2, and for
notational simplicity we have replaced $a_0$ by $A_0$ in eq.\ (\ref{<Z7>}). 
The quantity (\ref{<Z8>}) represents a generalized transverse 
projector \cite{<R120>}.

For $\Theta = 0$ eq.\ (\ref{<Z6>}) is precisely the path integral expression derived for the
Yang-Mills partition function in ref.\ \cite{<R108>} by explicitly
resolving Gauss' law in the modified axial gauge 
\begin{eqnarray}
\label{<Z9>}
A_0^{(\mathrm{ch})} = 0\,, \quad \partial_0 A_0^{(\mathrm{n})} = 0 \,,
\end{eqnarray}

where $A_0^{(\mathrm{n})}$ and $A_0^{(\mathrm{ch})}$ denote the ``neutral'' 
and ``charged'' parts of
$A_0$, which live in the
Cartan algebra $\mathcal{H}$ and in the coset $\mathcal{G/H}$, 
respectively.\footnote{More precisely, the
expression derived in reference \cite{<R108>} follows from eq.\ (\ref{<Z6>})
by a permutation of the Lorentz indices. In reference \cite{<R108>}, $A_0$
was integrated out to yield  Gauss' law and the gauge $A_3^{(\mathrm{ch})} = 0$ and
$\partial_3 A_3^{(\mathrm{n})} = 0$ was used to resolve Gauss' law.}
(See also ref.\ \cite{<R107>} where Gauss' law was resolved in the same gauge
in the canonical operator approach.\footnote{Similar gauges have been 
considered recently in refs.\ \cite{<R119A>}, \cite{<R119B>}.}) The gauge
(\ref{<Z9>}) is equivalent to the so-called Polyakov gauge defined by
diagonalizing the Polyakov line 
$\mathcal{P}\exp(-\int_0^T dx_0A_0)$.

In the standard Yang-Mills functional integral with gauge (\ref{<Z9>}) the reduced
Haar measure $\mathcal{D} \bar{\mu} (\exp (-Ta_0))$ of eq.\ (\ref{<Z6>}) arises from the
corresponding Faddeev-Popov determinant \cite{<R104>}. This connection has been recently
also established in $1+1$ dimensions \cite{<R119A>}. The 
derivation of the gauge fixed Yang-Mills partition function (\ref{<Z6>}) given
above, starting from the gauge invariant projection, is more concise than the
explicit resolution of Gauss' law in either the operator approach \cite{<R107>} or
in the standard functional integral approach \cite{<R108>}.
Of course both methods yield the same result.

\section{Magnetic monopoles and strings}
\label{sec5}

As we have seen in the previous section, the gauge invariant projection combined
with the diagonalization of the gauge function (\ref{<15>}) is equivalent to a
resolution of Gauss' law in the gauge (\ref{<Z9>}). This gauge represents a variant
of 't Hooft's Abelian gauges \cite{<R109>} in which magnetic monopoles are known
to occur. For later purposes it is instructive to briefly demonstrate the 
emergence of monopoles
in the diagonalization (\ref{<15>}), cf.\ also refs.\ \cite{<R109>}, \cite{<R110>},
\cite{<R114>}, \cite{<R126>}. For simplicity I consider again the gauge group
$SU (2)$.

Adopting the parameterization (\ref{<R18>}) chosen above for an element of the
gauge group, the coset matrix $V$ (\ref{<15>}) is defined by
\begin{eqnarray}
\label{<21>}
\hat{\bchi} \btau = V^\dagger \tau_3 V \,. 
\end{eqnarray}

Obviously the coset matrix $V$ depends only on the unit vector
$\hat{\bchi}(\theta, \phi)$, which can be parameterized in the usual
fashion by polar and
azimuthal angles $\theta$ and $\phi$. In the parameterization (\ref{<R18>}) the
element
of the invariant torus has the representation 
\begin{eqnarray}
\label{<22>}
\omega = e^{i \chi \tau_3} = \cos \chi + i \tau_3 \sin\chi \,.
\end{eqnarray}

Let us also quote the explicit representation of the coset matrix $V$
which diagonalizes the general group element. In fact, this matrix is
defined only up to an element of the Cartan subgroup $V \rightarrow g V$,  
$g \in H$, which does not change the decomposition (\ref{<15>}). Since this matrix
has to rotate an arbitrary vector in group space $\hat{\bchi}$ into the 
3-direction, this matrix can be chosen as the rotational matrix 
\begin{eqnarray}
\label{<28>}
V = e^{i \frac{\theta}{2} \mathbf{e}_\phi \mbox{\boldmath\scriptsize$\tau$}}\,, \quad 
\mathbf{e}_\phi = - \sin \phi\,
\mathbf{e}_1 + \cos \phi \,\mathbf{e}_2\,. 
\end{eqnarray}

For $\theta = \pi$ there is an ambiguity in the choice of the rotational axis
$\mathbf{e}_\phi$, when rotating $\hat{\bchi} (\theta = \pi) = 
- \mathbf{e}_3$ to
$\mathbf{e}_3$ (any
axis in the 1-2-plane can be chosen). This ambiguity is reflected by the
coordinate singularity of $\phi$ for $\theta = \pi$, which gives rise to a
line singularity of $V$ at $\theta = \pi$, where $V (\theta = \pi) = i
\mathbf{e}_\phi \btau$. For the identification of the singularities, 
the polar
coordinate representation (\ref{<28>})
is not very convenient since the polar coordinates themselves have singularities. To
exhibit the singular structure of $V$ it is more convenient to use Cartesian
coordinates, in which  the matrix $V$ can be chosen as
\begin{eqnarray}
\label{<29>}
V = - i \hat{\balpha} \btau\,, \quad \hat{\alpha}^a = 
\frac{\delta^{a3} +
\hat{\chi}^a}{\sqrt{2 (\hat{\chi}^3 + 1)}} \,, 
\quad \hat{\balpha}^2 = 1\,. 
\end{eqnarray}

Obviously this matrix has a string singularity at the negative 3-axis,
$\hat{\chi}^3 = -1$. The end points
of this line singularity correspond to the irregular group elements (\ref{<15>})
$\Omega = 1$ ($\chi = 2 k \pi$) and $\Omega = - 1$ ($\chi = (2 k + 1) 
\pi$).
At these points in group space, the corresponding induced gauge potential 
\begin{eqnarray}
\label{<31>}
\mathcal{A}_\alpha = V \partial_\alpha V^\dagger 
\end{eqnarray}

develops a magnetic monopole. The string singularity observed in the coset
matrix (\ref{<29>}) is nothing but the Dirac string (here in group space), which
interpolates between two monopoles with opposite magnetic charges or runs from a
monopole to infinity. Note that due to
the imposed boundary condition (\ref{<6A>}) 
$\lim_{r \to \infty} \chi (r) = n \pi$,
spatial infinity is mapped onto an irregular group element $\Omega = (- 1)^n$.
This implies that we may have a magnetic monopole at spatial
infinity and hence also a Dirac string which extends to
spatial infinity.

With the representation (\ref{<28>}), straightforward evaluation yields for the
induced gauge potential (\ref{<31>})
\begin{eqnarray}
\label{<30>}
\mathcal{A}_\alpha = - i \frac{1}{2} \mathbf{e}_\phi
\btau
\partial_\alpha \theta + \frac{i}{2} \sin \theta \mathbf{e}_\rho \cdot
\btau \, \partial_\alpha \phi 
+ i \frac{1}{2} (1 - \cos \theta) (\partial_\alpha \phi) \tau_3 \,, 
\end{eqnarray}

where we have used $\frac{\partial}{\partial \phi} \mathbf{e}_\phi = 
- \mathbf{e}_\rho$ and
$\mathbf{e}_\rho \times \mathbf{e}_\phi = \mathbf{e}_3$.
Here the last term in front of $\tau_3$ represents the gauge potential of a
magnetic monopole  with Dirac string at $\theta = \pi$.

\begin{figure}
\hspace{1cm}
\epsfig{file = 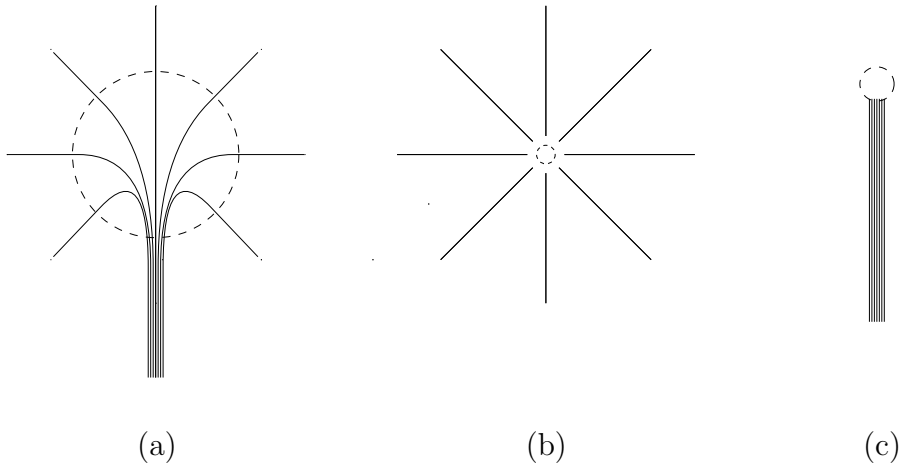, height = 5cm}\\[0.5cm] 
\hspace*{2.9cm}(a)\hspace{4.5cm} (b)\hspace{3.8cm} (c)
\caption{\label{fig2} Illustration of the various magnetic fields introduced in
sect.\ 5 for a
monopole configuration: (a) the Abelian magnetic field $\bcB$ (\ref{<E1>}),
(b) the non-Abelian contribution ($-\bsB$) (\ref{<E6>}) and (c) the total
magnetic field $\mathbf{B}^3$ (\ref{<E5>}). For illustrative purposes the
monopole fields have been
regularized. The true fields are obtained by contracting the dashed circle to a
point.}
\end{figure}

In fact, the Abelian magnetic field 
\begin{eqnarray}
\label{<E1>}
\bcB = \bnabla \times \bcA^3\,, \quad 
\bcA^3 = - 2 \mbox{Tr}(T^3 \bcA) 
\end{eqnarray}

is that of a Dirac monopole
shown in fig.\ \ref{fig2}a. The magnetic flux, flowing outward from the center 
of the
monopole, is oppositely the same as the magnetic flux of the Dirac string,
flowing inside the center, and the net magnetic flux of the monopole and
the Dirac string vanishes,
\begin{eqnarray}
\label{<E3>}
\int_{S_\infty^2} d \mathbf{\Sigma} \bcB = 0 \,, 
\end{eqnarray}

where $S_\infty^2$ denotes the surface of $R^3$.

Since the induced gauge potential (\ref{<31>}) is pure gauge, its total
(non-Abelian) field strength
\begin{eqnarray}
\label{<E4>}
F_{\mu \nu} [\mathcal{A}] = \partial_\mu \mathcal{A}_\nu - \partial_\nu 
\mathcal{A}_\mu + 
[\mathcal{A}_\mu, \mathcal{A}_\nu] 
\end{eqnarray}

vanishes except at the singularities of $V$, where the r.h.s.\ can be 
non-zero. In the presence of a magnetic monopole, $V$ has a string
singularity and we expect a non-vanishing total magnetic field 
\begin{eqnarray}
\label{<E5>}
B_i [\mathcal{A}] =  \frac{1}{2} \epsilon_{ijk} F_{jk} [\mathcal{A}]  
\end{eqnarray}

at the Dirac string. These are the magnetic strings found in 
ref.\ \cite{<R114>}, where it was observed that unlike the Abelian 
field (\ref{<E1>}),
the total non-Abelian field (\ref{<E4>}) does not contain monopole type fields,
but
only strings of magnetic flux, which are the remnants of the Dirac string of the
Abelian magnetic field. What happens is that, in the total field strength
(\ref{<E4>}), the non-Abelian commutator term 
$[\mathcal{A}_\mu, \mathcal{A}_\nu]$ 
cancels the monopole part of the Abelian magnetic field (\ref{<E1>}), leaving
only the Dirac
string. Therefore, the magnetic field defined by 
\begin{eqnarray}
\label{<E6>}
\mathsf{B}_i = \mathcal{B}_i - B_i^3 = - \frac{1}{2} \epsilon_{ijk} 
[\mathcal{A}_j, \mathcal{A}_k]^3 = \epsilon_{ijk} \mbox{Tr} ([\mathcal{A}_j, 
\mathcal{A}_k]T_3) 
\end{eqnarray}

represents the magnetic field of a monopole without the Dirac string. This field
then obviously satisfies 
\begin{eqnarray}
\label{<E7>}
\bnabla \bsB = 4 \pi \sum_i m_i \delta^{(3)} (\mathbf{x} - \bar{\mathbf{x}}_i)\,,
\end{eqnarray}

where $\bar{\mathbf{x}}_i$ are the positions of the monopoles and $m_i$ their
magnetic charges, which agree with the total magnetic flux of the monopole
field,
\begin{eqnarray}
\label{<E8>}
m_i = \frac{1}{4 \pi} \int_{S_\varepsilon^2(i)} d \bsigma
\bsB = \frac{1}{4 \pi} \int d {\sigma}_i \epsilon_{ijk} \mbox{Tr} 
([\mathcal{A}_j, \mathcal{A}_k] T_3)\,, 
\end{eqnarray}

where $S_\varepsilon^2(i)$ is a sphere of infinitesimal radius around
the center of the monopole, $\bar{\mathbf{x}}_i$. It is straightforward to show that the
magnetic flux (\ref{<E8>}) agrees with the winding number $m [\hat
{\bchi}] \in \Pi_2 (SU (2) / U (1) ) = \Pi_2 (S_2)$ of the mapping 
$\hat{\bchi} (\hat{\mathbf{x}})$ given in eq.\ (\ref{<R27>}), cf.\ also 
refs.\ \cite{<R110>}, \cite{<R114>}.  
This ensures that the charge of the magnetic monopole is quantized
\begin{eqnarray}
\label{<E11>}
m_i = 0, \pm 1, \pm 2, \dots \,.
\end{eqnarray}

The $\bsB$ field obviously agrees with the Abelian magnetic field $\bcB$
everywhere except at the Dirac string. Therefore, the magnetic flux
(\ref{<E8>}) can be alternatively evaluated from the Abelian field $\bcB$
by leaving out that point $\bar{\bar{\mathbf{x}}}$ of $S_\infty^2$ where the Dirac 
string pierces the surface,
\begin{eqnarray}
\label{<E12>}
m_i = \frac{1}{4 \pi} \int\limits_{\tilde{S}^2(i)} d \bsigma \cdot 
\bcB\,.
\end{eqnarray}

The integration is then over a punctured sphere $\tilde{S}^2(i) =
{S}_{\varepsilon}^2(i) \setminus \{ \bar{\bar{\mathbf{x}}}_i \}$ and can be 
performed by applying Stoke's theorem, 
\begin{eqnarray}
\label{<E13>}
m_i = \frac{1}{4 \pi} \int\limits_{C_i} d \mathbf{x}\, \bcA^3 \,, 
\end{eqnarray}

where $C_i$ is an infinitesimal circle enclosing the Dirac string of the
monopole. 
Inserting here the explicit form of $\bcA^3$ (\ref{<30>}) and taking into
account that on the infinitesimal circle $C_i$ around the Dirac string we have
$\theta\simeq\pi$ we obtain for the magnetic flux (\ref{<E13>}) 
\begin{eqnarray}
m_i=-\frac{1}{2\pi}\int\limits_{C_i}d\mathbf{x}\,\bnabla\phi=-m[\phi]\,.
\end{eqnarray}

This is the winding number $m[\phi]\in\Pi_1(U(1))$ of the mapping
\begin{eqnarray}
\phi(x): \mathbf{x}\in C_i \simeq S_1 \to \phi \in S_1\simeq U(1)\,.
\end{eqnarray}

As discussed above the magnetic flux $m_i$ is also given by the winding number
(\ref{<R27>}) $m[\hat{\bchi}]\in\Pi_2(SU(2)/U(1))$. The equality of both winding
numbers, $m[\phi]$ and $m[\hat{\bchi}]$, is guaranteed by the relation
\begin{eqnarray}
\Pi_2(SU(2)/U(1))=\Pi_1(U(1))\,.
\end{eqnarray}

In the next section we will see that it is the flux (\ref{<E8>}) 
of the monopole field $\bsB$ which also determines the winding number 
$n[\Omega]\in \Pi_3(SU(2))$.

Let us explicitly quote the various magnetic
fields introduced above for a generic mapping $\Omega (\mathbf{x})$ given by 
the generalized hedgehog field\footnote{Recall, while $(\vartheta,\varphi)$ are
polar and azimuthal angle in ordinary space, $(\theta,\phi)$ denote the 
corresponding quantities in group space.}
\begin{eqnarray}
&&\chi=\chi(r) \quad\mbox{with}\quad \chi(0)=0\,,\quad \chi(\infty)=n\pi
\label{neu2}\\
&&\theta=p\vartheta\,,\quad \phi=q\varphi\,.\nonumber
\end{eqnarray}

To get a handle on the singularities of $\mathcal{A}_\alpha$ (\ref{<31>}) we use
the representation (\ref{<29>}) for $V$ and introduce a regularization
\cite{<R114>}
\begin{eqnarray}
\mathcal{A}_k^a=-\lim\limits_{\varepsilon\to
0}\,\frac{2}{\alpha^2+\varepsilon^2}\,\epsilon^{abc}\alpha^b\partial_k\alpha^c
\end{eqnarray}

where
\begin{eqnarray}
\alpha^a=\chi^a+\chi\,\delta^{a3}\,,\quad \alpha^2=2\,\chi^2(1+\hat{\chi}^3)\,.
\end{eqnarray}

For $\chi=\chi(r)$ independent of $\vartheta$ and $\varphi$ one finds in
spherical coordinates
\begin{eqnarray}
\mathcal{A}_r^3&=&0 \nonumber\\
\mathcal{A}_\vartheta^3&=&-\lim\limits_{\varepsilon\to
0}\,\frac{2\chi^2}{\alpha^2+\varepsilon^2}\,\epsilon^{3bc}\,\hat{\chi}^b\,
\frac{1}{r}\frac{\partial}{\partial\vartheta}\hat{\chi}^c \\
\mathcal{A}_\varphi^3&=&-\lim\limits_{\varepsilon\to
0}\,\frac{2\chi^2}{\alpha^2+\varepsilon^2}\,\epsilon^{3bc}\,\hat{\chi}^b\,
\frac{1}{r\sin\vartheta}\frac{\partial}{\partial\varphi}\hat{\chi}^c \nonumber
\end{eqnarray}

which gives rise to the Abelian magnetic field (\ref{<E1>})
\begin{eqnarray}
\bcB=-q\,\frac{\sin p\vartheta}{r^2\sin\vartheta}\,\lim\limits_{\varepsilon\to
0}\left\{p\,\frac{\alpha^4}{(\alpha^2+\varepsilon^2)^2}\,\mathbf{e}_r+
\frac{\varepsilon^2}{(\alpha^2+\varepsilon^2)^2}\,4\chi^2\left[p\cos
p\vartheta\,
\mathbf{e}_r -\frac{r\chi'}{\chi}\sin p\vartheta\, \mathbf{e}_\vartheta\right]
\right\}
\end{eqnarray}

with $\chi'=d\chi/dr$.
The first term is regular for $\varepsilon\to 0$. Using
\begin{eqnarray}
\lim\limits_{\varepsilon\to 0}\,\frac{\varepsilon^2}{(\alpha^2+\varepsilon^2)^2}=
\pi\delta(\chi^1)\delta(\chi^2)\Theta(-\chi^3)
\end{eqnarray}

we obtain for $\varepsilon\to 0$
\begin{eqnarray}
\bcB=-q\,\frac{\sin p\vartheta}{r^2\sin\vartheta}\left\{p\,\mathbf{e}_r
+\pi\delta(\chi^1)\delta(\chi^2)\Theta(-\chi^3)\,4\chi^2\left[p\cos p\vartheta\,
\mathbf{e}_r -\frac{r\chi'}{\chi}\sin p\vartheta\, \mathbf{e}_\vartheta\right]
\right\}\,.
\label{neu1}
\end{eqnarray}

For simplicity let us consider the case $p=q=1$.
Near the singularity (monopole position) $\chi=0$ we have 
$\chi(r) \simeq r\chi'(0)$ so
that $r\chi'/\chi \simeq 1$ and the magnetic field (\ref{neu1}) reduces to
($\mathbf{e}_3=\cos\vartheta\mathbf{e}_r-\sin\vartheta\mathbf{e}_\vartheta$) 
\begin{eqnarray}
\bcB = -\frac{\hat{\mathbf{x}}}{r^2} - 4 \pi \Theta (-x_3) \delta(x_1) \delta
(x_2) \mathbf{e}_3\,,
\end{eqnarray}

which is the magnetic field of the familiar Dirac monopole with magnetic charge
$m = 1$, in agreement with the topological quantization of the magnetic
flux. For the field $\bsB$ (\ref{<E6>}) one finds near the singularity
\begin{eqnarray}
\bsB =-\frac{\hat{\mathbf{x}}}{r^2} \,, 
\end{eqnarray}

which is a monopole field without the Dirac string.

Before concluding this section let us comment on the physical meaning of the
induced gauge potential $\mathcal{A}_\alpha$ (\ref{<31>}) and, in particular, of
the monopole singularities: The (time-independent) gauge function
$\Omega(\mathbf{x})=\exp(-TA_0(\mathbf{x}))$ is nothing but the Polyakov line in
the gauge $\partial_0A_0=0$ and diagonalization of this map
$\Omega(\mathbf{x})=V^\dagger\omega V$ is equivalent to going to the Polyakov
gauge. Lattice calculations performed in the Polyakov gauge show not only
Abelian dominance but also dominance of magnetic monopoles \cite{XX}, i.e.\ most
of the string tension comes from the magnetic monopole configurations alone. In
this respect the magnetic monopoles arising in the induced gauge potential
$\mathcal{A}_\alpha=V\partial_\alpha V^\dagger$ from singular gauge
transformations $V$ (\ref{<28>}) represent the dominant infrared degrees of
freedom. This is very analogous to Yang-Mills theory in the maximum Abelian
gauge defined by
\begin{eqnarray}
\left[\partial_\mu+A_\mu^{(\mathrm{n})},A_\mu^{(\mathrm{ch})}\right]=0
\end{eqnarray}

where the infrared dominance of magnetic monopoles is even more pronounced
\cite{<R111>}.

\section{Relation between winding number and monopole charges}

In what follows, we consider how the winding number (\ref{<6>}) transforms
under the Cartan decomposition (\ref{<15>}). Thereby we will derive a 
relation between the winding number $n
[\Omega = V^\dagger \omega V]$ and the magnetic charges of the monopoles induced by the
(coset) gauge transformation $V (x)$.\footnote{ A naive application of 
eq.\ (\ref{<R14>})
would yield the result $n [\Omega] = n [V^\dagger] + n [\omega] + n [V] = 0$ since $n
[V^\dagger] = - n [V]$ and $n [\omega] = 0$. However, unlike $\Omega$, $V$ does not
approach an angle-independent limit for $r \to \infty$, which is required for 
a gauge
function in order for its winding number to be well defined.}
                                        
The relevant current can be rewritten as
\begin{eqnarray}
\label{<17>}
L_\alpha = \Omega \partial_\alpha \Omega^\dagger & = & V^\dagger \omega
\mathcal{A}_\alpha \omega^\dagger V + V^\dagger
s_\alpha V + V^\dagger \partial_\alpha V \nonumber\\
& = & V^\dagger (\tilde{\mathcal{A}}_\alpha - \mathcal{A}_\alpha + s_\alpha) V
\,.
\end{eqnarray}

Here we have used the definition of $\mathcal{A}$ (\ref{<31>}) and the abbreviations
\begin{eqnarray}
\label{<18>}
\tilde{\mathcal{A}}_\alpha =
 \omega \mathcal{A}_\alpha
 \omega^\dagger\,,
\quad s_\alpha = \omega \partial_\alpha \omega^\dagger \,.
\end{eqnarray}

Using the cyclic properties of the trace and the fact that two elements of the
Cartan algebra commute, i.e.\ $[s_\alpha, s_\beta] = 0$, the expression for the
winding number (\ref{<6>}) can be reduced to
\begin{eqnarray}
\label{<19>}
n [\Omega] = \frac{1}{24 \pi^2} \int d^3 x \,
\epsilon^{\alpha \beta \gamma}
\mbox{Tr} \Big\{&&\hspace{-0.7cm}(\tilde{\mathcal{A}}_\alpha - \mathcal{A}_\alpha)
(\tilde{\mathcal{A}}_\beta - 
\mathcal{A}_\beta) (\tilde{\mathcal{A}}_\gamma - \mathcal{A}_\gamma) \nonumber\\
&+& 3 (\tilde{\mathcal{A}}_\alpha - \mathcal{A}_\alpha) 
(\tilde{\mathcal{A}}_\beta - \mathcal{A}_\beta)
 s_\gamma \quad\Big\}\,.
\end{eqnarray}

Using the Cartan decomposition of the Lie algebra of the gauge group, it is
straightforward to show that the quantity $\tilde{\mathcal{A}}_\alpha - 
\mathcal{A}_\alpha$ lives entirely in the coset space ${\cal G} / {\cal H}$. 
Therefore,
the first term in eq.\ (\ref{<19>}) vanishes for $SU (2)$. 
Indeed, with the above adopted
parameterization of the gauge
group (\ref{<19>}), (\ref{<22>}), one finds 
\begin{eqnarray}
\label{<23>}
s_\alpha &=& i \tau_3 \partial_\alpha \chi\,,  \\
\label{<24>}
\tilde{\mathcal{A}}_\alpha - \mathcal{A}_\alpha &=& - 2 \sin^2 \chi
\mathcal{A}_\alpha^{(\mathrm{ch})} - \sin 2 \chi
\epsilon_{3 \bar{a} \bar{b}} \, \mathcal{A}_\alpha^{\bar{a}} 
\, T^{\bar{b}} \,, 
\end{eqnarray}

where $\mathcal{A}_\alpha^{(\mathrm{ch})} = \mathcal{A}_\alpha^{\bar{a}} 
T^{\bar {a}}$ denotes the charged
part of $\mathcal{A}_\alpha$. (The indices $\bar{a}, \bar{b} = 1, 2$ are restricted
to the generators of the coset ${\cal G} / {\cal H}$.) 
The winding number (\ref{<19>}) then becomes 
\begin{eqnarray}
\label{<27>}
n [\Omega] = \frac{i}{4 \pi^2} \int d^3 x \,\sin ^2 \chi 
\epsilon^{\alpha \beta \gamma} \partial_\gamma \chi \mbox{Tr}
([\mathcal{A}_\alpha, \mathcal{A}_\beta] \tau_3)\,. 
\end{eqnarray}

Using 
\begin{eqnarray}
\sin^2 \chi \partial_k \chi = \frac{1}{2} \partial_k (\chi - \sin
\chi \cos \chi)
\end{eqnarray}

and the definition of the monopole field $\bsB$ (\ref{<E6>}), the topological
charge can be expressed as 
\begin{eqnarray}
n [\Omega] = - \frac{1}{4 \pi^2} \int d^3 \Sigma_0 \mathsf{B}_k 
\partial_k (\chi - \sin \chi \cos \chi)\,.
\end{eqnarray}

Performing a partial integration and using Gauss' theorem, we obtain 
\begin{eqnarray}
\label{<88>}
n [\Omega]  & = & - \frac{1}{4 \pi^2} \int d^2 \sigma_k \mathsf{B}_k
(\chi - \sin \chi \cos \chi) \nonumber\\
&& +  \frac{1}{4 \pi^2} \int d^3 \Sigma_0 (\chi - \sin\chi \cos\chi)
\bnabla \bsB \nonumber\\
& = & n^{(1)} + n^{(2)} \,.
\end{eqnarray}

\begin{figure}
\begin{center}
\epsfig{file = 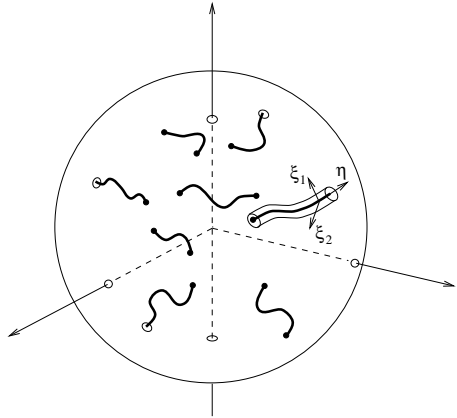, width = 6cm}
\end{center}
\caption{\label{fig3} Illustration of a gas of magnetic strings. For one
of the strings the
local coordinate system ($\eta, \xi_1, \xi_2$) (see text) has been indicated by
the tube.}
\end{figure}

Due to our chosen boundary condition $\lim_{r \to \infty} \chi (r) = n \pi$,
the first term yields 
\begin{eqnarray}
n^{(1)} = - \frac{n}{4 \pi} \int d \bsigma \cdot \bsB
 = - n m
[\hat{\bchi}] \,,
\end{eqnarray}

where
\begin{eqnarray}
m [\hat{\bchi}] = \sum_i m_i [\hat{\bchi}]
\end{eqnarray}

is the magnetic flux of the monopoles (\ref{<E8>}), (\ref{<R27>}). 
The second term $n^{(2)}$ receives
contributions from the monopoles, which are the sources of the field
$\bsB$. Using eq.\ (\ref{<E7>}), the integration in $n^{(2)}$ can performed
and we obtain 
\begin{eqnarray}
n^{(2)} = \frac{1}{\pi} \sum_k (\chi (\bar{x}_k) - \sin\chi 
(\bar{\mathbf{x}}_k) \cos\chi (\bar{\mathbf{x}}_k)) m_k \,,
\end{eqnarray}

where $\bar{\mathbf{x}}_k$ denotes the positions of the magnetic monopoles. At a
monopole position, where the gauge function $\Omega (\mathbf{x})$ is irregular, 
we have (cf.\ eq.\ (\ref{<R18>}))
\begin{eqnarray}
\chi (\bar{\mathbf{x}}_k) = n_k \pi \,,
\end{eqnarray}

where $n_k$ is integer. Hence we find 
\begin{eqnarray}
n^{(2)} = \sum_k n_k m_k
\end{eqnarray}

and the total winding number (\ref{<88>}) becomes 
\begin{eqnarray}
\label{<93>}
n [\Omega] = - \sum_k (n - n_k) m_k = \sum_k \ell_k m_k \,,
\end{eqnarray}

where 
\begin{eqnarray}
\label{<94>}
\ell_k = n_k - n 
\end{eqnarray}

times $\pi$ are the (signed) lengths of the Dirac strings of the monopoles in 
group
space, cf.\ fig.\ \ref{fig3} and the appendix.
Note that for gauge functions $\Omega (\mathbf{x})$ with well-defined 
winding number $n[\Omega]$ (\ref{<6>}), the $n_k$ and hence the $\ell_k$ are 
integer. 

Eq.\ (\ref{<93>}) expresses the winding number of the mapping $n [\Omega = V \omega
V^\dagger]$ in terms of the charges $m_k$ of the magnetic monopoles of the induced
gauge potential ${\cal A}_i = V \partial_i V^\dagger$ and the properties 
$\ell_k$ of the Cartan element $\omega$. Let us emphasize that $\ell_k$ are the
(integer valued) lengths of the string singularities of $\mathcal{A}_i$ in 
group
space. In ordinary space the mapping of such a string singularity (which is
usually referred to as the Dirac string) can have arbitrary length and furthermore
the string singularity can even be split into several strings or deformed to conic sheet
singularities. In the appendix we will illustrate this by means of a few
generic mappings $\Omega (\mathbf{x})$ which illustrate what happens in the general
case.

A more physical derivation of the above result can be obtained in the following
way. The Pontryagin index of the gauge field entering
the Yang-Mills partition function (\ref{<Z3>}) is given by 
\begin{eqnarray}
\label{H1}
\nu [A'] \equiv \nu [a_0, A_i^V] = - n\,. 
\end{eqnarray}

This can be easily seen by noticing that the gauge field in eq.\ (\ref{<Z3>})
differs from the one in (\ref{<14>}), $A_\mu^U$, by the time-independent gauge
transformation $V \in G/H$, which in view of $\Pi_3 (SU (2)/U(1)) =1$
does not change the Pontryagin index. The relation (\ref{H1}) follows then from
eqs.\ (\ref{<R15>}) and (\ref{<R15A>}). On the other hand, the Pontryagin index
can be expressed in terms of the colour electric and magnetic fields by
\begin{eqnarray}
\label{H2}
\nu [A] = \frac{1}{4 \pi^2} \int dt \int d^3 x\, \mbox{Tr} (\mathbf{E} 
\mathbf{B})\,. 
\end{eqnarray}

Obviously only long range fields of the monopole type give here non-zero
contributions, cf.\ eq.\ (\ref{<R4>}). For the calculation of the Pontryagin index it is therefore
sufficient to keep from the spatial gauge field only the induced gauge field
(\ref{<31>}), which contains all the monopoles and by construction is
time-independent. The electric field is then given by
\begin{eqnarray}
\label{H3}
E_i = - \partial_i a_0 + [a_0,\mathcal{A}_i] \,,
\end{eqnarray}

where the Abelian field $a_0$ is defined by eq.\ (\ref{<Z4>}). From the induced
gauge field $\mathcal{A}_i = V \partial_i V^\dagger$ only the Abelian monopole part
is long range, cf.\ eq.\ (\ref{<30>}), and has hence to be included. Then the
commutator term in eq.\ (\ref{H3}) vanishes and we obtain with 
eqs.\ (\ref{<Z4>}), (\ref{<22>}) 
\begin{eqnarray}
\label{H4}
E_i = \frac{1}{T} \partial_i \ln \omega_n = - \frac{2}{T} T_3 \partial_i \chi
\,.
\end{eqnarray}

Obviously the quantity $\chi$ figures as the scalar potential for the Abelian
electric field. Inserting eq.\ (\ref{H4}) into eq.\ (\ref{H2}), the Pontryagin
index becomes 
\begin{eqnarray}
\label{H5}
\nu [A'] = \nu [a_0, \mathcal{A}_i] = - \frac{1}{4 \pi^2} \cdot \frac{1}{T}
\int dt \int d^3 x \,(\partial_k \chi) B_k^3 \,,  
\end{eqnarray}
 
where $B_k^3 =  - 2 \mbox{Tr}(T_3 B_k [\mathcal{A}])$ is the Abelian component of the full
(non-Abelian) magnetic field. As shown in sect.\ 5, this field vanishes, since
$\mathcal{A}_i$ is pure gauge, except for the string singularities. Therefore the
Pontryagin index receives contributions only from the Dirac strings of the
magnetic monopoles. It becomes hence a sum of the contributions  from the
individual monopoles,
\begin{eqnarray}
\label{H6}
\nu [A'] = \sum_i \nu_{(i)} [A']\,, 
\end{eqnarray}

where $\nu_{(i)}$ is the contribution of the $i$th monopole. To calculate this
contribution it is convenient to introduce local coordinates in the
neighborhood of the string, see fig.\ \ref{fig3}. Let $\eta$ and $\xi_1, \xi_2$ denote
the coordinates parallel and orthogonal, respectively, to the Dirac string. The
magnetic flux is conserved (in its magnitude) along the Dirac string and hence
independent of $\eta$. On the other hand, $\chi$ depends only on $\eta$.
Hence the integration orthogonal to the string can be explicitly performed and
yields ($\int d\xi_1\, d\xi_2\,(\mathbf{B}^3\cdot\mathbf{e}_\eta)=4\pi m_i$)
\begin{eqnarray}
\label{H7}
\nu_{(i)} [A'] = m_i \frac{1}{\pi} \int d \eta \frac{\partial \chi}{\partial 
\eta} \,,
\end{eqnarray}

where $(- m_i)$ is the magnetic flux of the Dirac string, which is equal in
magnitude but opposite in sign to the magnetic flux of the spherically symmetric
monopole field $\bsB$, defined by eq.\ (\ref{<E6>}). In view of our boundary
condition (\ref{<R25>}), the Dirac string runs between the singular field
configuration $\chi (\eta = 0) = n_i \pi$ ($\eta = 0$ corresponds here to the
monopole position while $\eta \to \infty$ corresponds to spatial infinity) and 
$\chi (\infty) = n \pi$. Hence, with eq.\ (\ref{<94>}),
we obtain 
\begin{eqnarray}
\nu_{(i)} [A'] = - \ell_i m_i \,,
\label{<102>}
\end{eqnarray}

which together with eqs.\ (\ref{H6}) and (\ref{H1}) agrees with our previous
result (\ref{<93>}).

The result obtained above shows that magnetic monopoles with Dirac strings running
to infinity are sufficient to account for the topologically non-trivial sectors of
Yang-Mills theory. While magnetic monopoles connected by finite strings do not
contribute to the topology, they are presumably the relevant infrared degrees of
freedom which are responsible for confinement \cite{<R111>}, \cite{<RR>}.

Finally, let me comment on previous related work. In ref.\ \cite{<R121>} the
Pontryagin index of dyon configurations, treated as periodic in time, was shown
to coincide with the magnetic charge. This corresponds to the case $\ell_i=1$ in
eq.\ (\ref{<102>}). There have been also recent investigations of 
Yang-Mills theory in the modified axial gauge (\ref{<Z9>}) on the torus 
\cite{<R122>}, where large gauge transformations have been found to induce 
magnetic flux.

\section{Concluding remarks}

In this paper I have demonstrated that gauge invariant projection provides an 
efficient way of resolving Gauss' law in Yang-Mills theory. For the 
partition function, the projection onto gauge invariant orbits amounts to 
integration over the Cartan  subgroup. The irregular group elements 
give rise to magnetic monopoles which are responsible for the non-trivial
topological structure of the gauge fields. 
I have explicitly shown how these magnetic charges build up the Pontryagin index.
The latter is determined by the magnetic charges and the lengths of the 
associated Dirac strings in group space (cf.\ eq.\ (\ref{<93>})).

An important conclusion of the present paper is that the magnetic monopoles are
entirely sufficient to account for the non-trivial topological structure of
Yang-Mills theory. This result is perhaps not surprising since
intuitively one may expect that the global, i.e.\ topological properties 
of gauge fields are related to the long range monopole fields. Furthermore this
result is also consistent with the monopole dominance seen in lattice
calculations \cite{<R111>}, \cite{XX}.

\subsubsection*{Acknowledgements:}

Discussions with M.\ Engelhardt, H.\ Griesshammer, K.\ Langfeld, M.\ Quandt 
and H.\ Weigel are gratefully acknowledged. The author is also grateful to 
A.\ Schaefke for her kind assistance in preparing the \TeX-file, and in
particular the figures. 

\newpage
\appendix

\section{Generic mappings}

By definition of homotopy classes, any mapping $\Omega (\mathbf{x}) \in SU (2)$
can
be smoothly deformed into a field of the hedgehog type (with the same winding
number). In this sense,
the hedgehog is a generic topologically non-trivial mapping.

As a first illustrative example, we consider the ordinary hedgehog with winding
number $n [\Omega] = 1$,
defined by eq.\ (\ref{<R18>}) with
\begin{eqnarray}
\label{<33>}
\hat{\bchi} = \hat{\mathbf{x}}\,, \quad \chi (\mathbf{x}) = 
\chi (| \mathbf{x} | )
\end{eqnarray}

and with the boundary conditions 
\begin{eqnarray}
\label{<34>}
\chi (0) = 0\,, \quad \chi (\infty) = \pi\,. 
\end{eqnarray}

The field obviously has the right asymptotics (\ref{<6A>}) and yields a
conformal
mapping of ordinary space onto group space. Since $\hat{\bchi} =
\hat{\mathbf{x}}$, the Dirac string is obviously now also along the 
3-axis in ordinary space and from (\ref{<34>}) it follows that its length (in
group space) is
$\ell=1$, cf.\ eq.\ (\ref{<93>}).

From the hedgehog with winding number
one, we can construct the hedgehog with winding number $n$ by taking the hedgehog
field to the $n$th power. The new mapping still satisfies eq.\ (\ref{<33>}) and is
hence diagonalized by the same $V (\mathbf{x})$, but the boundary condition
(\ref{<34>})
is changed to 
\begin{eqnarray}
\chi (0) = 0\,, \qquad \chi (\infty) = n \pi \,,
\end{eqnarray}

which increases the length $\ell$ of the Dirac string in group space by a factor $n$ but
leaves its position unchanged. Also the induced monopole at $\mathbf{x} = 0$ has the same
magnetic charge $m = 1$ as in the case $n = 1$. Thus the increase in the winding
number by going from $\Omega$ to $\Omega^n$ is entirely due to the increase of the length
of the Dirac string in group space.

There are, however, alternative modifications of the
hedgehog which lead to higher winding numbers. Consider the mapping 
\begin{eqnarray}
\label{<38>}
\theta = \vartheta\,, \quad \phi = n \varphi \,,
\end{eqnarray}

where $\theta, \phi$ are the polar and azimuthal angle in group space, while
$\vartheta, \varphi$ are the corresponding quantities in ordinary space.
Furthermore, we assume that $\chi (r)$ depends only on the radius and
satisfies the boundary conditions (\ref{<34>}). 
Obviously this field has again winding number $n$. Since this map represents the
identity map for the polar angle $\theta = \vartheta$, there is only one Dirac
string along the
negative 3-axis, like in group space. However, this string now carries
$n$ times the magnetic flux of the hedgehog field with winding number one, as
can
be easily shown by using Stoke's theorem. Obviously, in this field configuration
$n$ monopoles are sitting on top of each other in the origin of the coordinate
system. From these $n$ monopoles, magnetic flux lines run to infinity, where
they are contracted by oppositely charged monopoles. 

Finally let us consider the mapping 
\begin{eqnarray}
\label{<36>}
\theta = p \vartheta\,, \quad \phi = \varphi\,, 
\end{eqnarray}

Naively one would expect that this mapping $\Omega(\mathbf{x})$ has also winding number
$n[\Omega]=p$. This is, however, not true. To see this let us explicitly
calculate the winding number for the generalized hedgehog mapping defined by
eq.\ (\ref{neu2}). The above considered examples are special cases of this
mapping.

Using eq.\ (\ref{neu3}) the winding number (\ref{<6>}) is expressed as
\begin{eqnarray}
n[\Omega]=\frac{1}{24\pi^2}\int d^3x\,
\epsilon_{ijk}\mbox{Tr}\,(L_i\partial_jL_k)
\end{eqnarray} 

where in the representation (\ref{<R18>})
\begin{eqnarray}
L_k=-i\hat{\bchi}\btau\partial_k\chi -\frac{i}{2}\sin 2\chi\,
(\partial_k\hat{\bchi})\btau+i\sin^2\chi\,(\hat{\bchi}\times
\partial_k\hat{\bchi})\cdot \btau
\end{eqnarray} 

so that
\begin{eqnarray}
&&\epsilon_{ijk}\mbox{Tr}\,(L_i\partial_jL_k)=2\epsilon_{ijk}\left\{3\sin^2\chi\,
(\partial_i\chi)(\partial_j\hat{\bchi}\times
\partial_k\hat{\bchi})\cdot\hat{\bchi}\,+\right.
\label{neu4}\\
&&\hspace*{7cm}\left.+\cos\chi\sin^3\chi\,
\partial_i\hat{\bchi}\cdot(\partial_j\hat{\bchi}\times
\partial_k\hat{\bchi}) \right\}\,.
\nonumber
\end{eqnarray}

For the mapping (\ref{neu2}) it is convenient to switch to spherical
coordinates. Then it is seen that the last term on the r.h.s.\  of eq.\
(\ref{neu4}) vanishes and the winding number becomes
\begin{eqnarray}
n[\Omega]=n\, m[\hat{\bchi}]\,.
\end{eqnarray}

Here we have used
\begin{eqnarray}
\int\limits_0^\infty dr\,
\chi'(r)\sin^2\chi(r)=\int\limits_{\chi(0)}^{\chi(\infty)}d\chi\,\sin^2\chi
=\frac{n\pi}{2}
\end{eqnarray} 

and
\begin{eqnarray}
m[\hat{\bchi}]=\frac{1}{4\pi}\int\limits_0^\pi
d\vartheta\int\limits_0^{2\pi}d\varphi\,
\left(\frac{\partial\hat{\bchi}}{\partial\vartheta}\times
\frac{\partial\hat{\bchi}}{\partial\varphi}\right)\cdot\hat{\bchi}
\end{eqnarray}  

is the $\Pi_2(S_2)$ winding number (\ref{<R27>}) of the mapping
$\hat{\bchi}(\vartheta,\varphi)$ which, as shown in section \ref{sec5}, 
coincides with the magnetic charge. For the mapping (\ref{neu2}) straightforward
evaluation yields
\begin{eqnarray}
m[\hat{\bchi}]=-q\frac{1}{2}(1-(-1)^p)\,.
\end{eqnarray}

Hence the magnetic charge and consequently also the winding number $n[\Omega]$
vanishes for even $p$. This result can be easily understood by noticing that
$\theta=\pi+\alpha$ is equivalent to $\theta=\pi-\alpha$. Increasing $\theta$
from $\theta=0$ to $\theta=\pi$ the $SU(2)$ group is covered once, by increasing
$\theta$ further to $\theta=2\pi$ the group is again uncovered.

Let us now return to the discussion of the mapping (\ref{<36>}), i.e.\ $n=1$,
$q=1$ in eq.\ (\ref{neu2}), and let us also determine the Dirac string in 
ordinary
space, situated still on the negative 3-axis in group space, which occurs
 for 
$\theta = (2 k + 1) \pi$, $k = 0, 1, 2,\dots$, cf.\ eqs.\ (\ref{<28>}), (\ref{<30>}).
Since $0 \le \vartheta \le \pi$ the
integer $k$ is restricted to 
\begin{eqnarray}
\label{<37>}
k \le \frac{p - 1}{2}\,. 
\end{eqnarray}

\begin{figure}
\begin{minipage}{5cm}
\begin{center}
\epsfig{file = 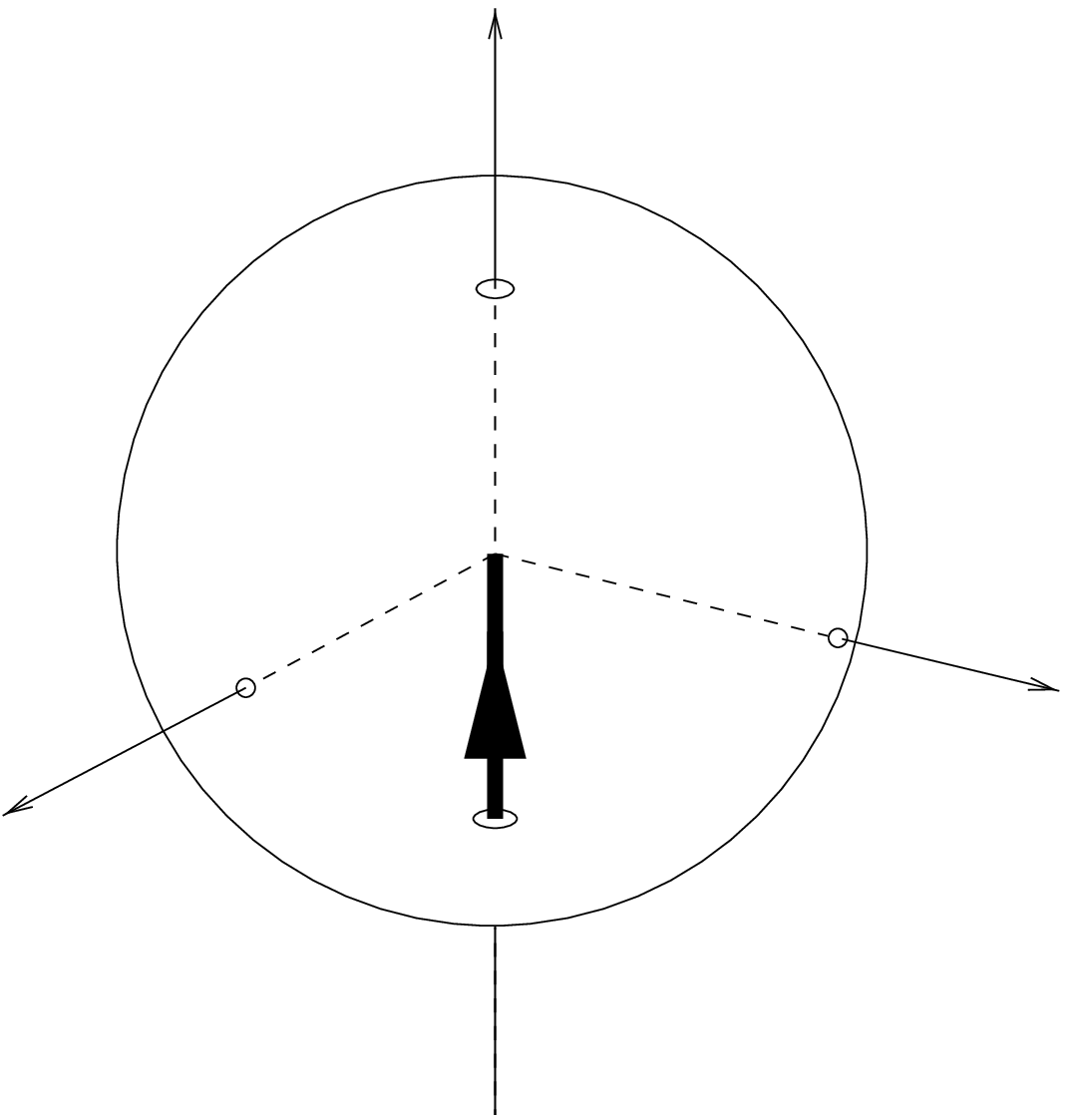, width = 5cm} \\
(a) 
\end{center}
\end{minipage}
\hfill
\begin{minipage}{5cm}
\begin{center}
\epsfig{file = 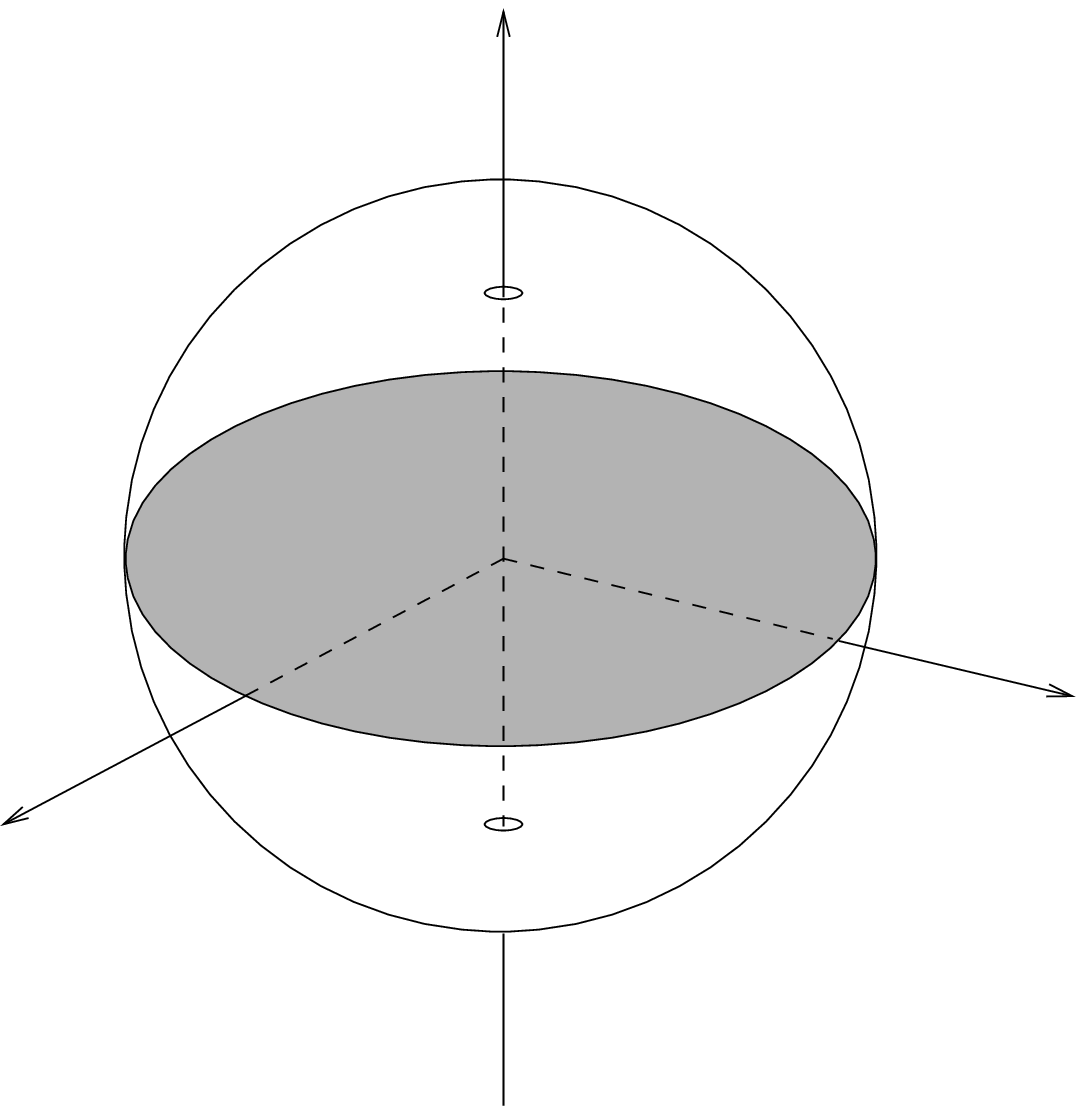, width = 5cm} \\
(b) 
\end{center}
\end{minipage}
\hfill
\begin{minipage}{5cm}
\begin{center}
\epsfig{file = 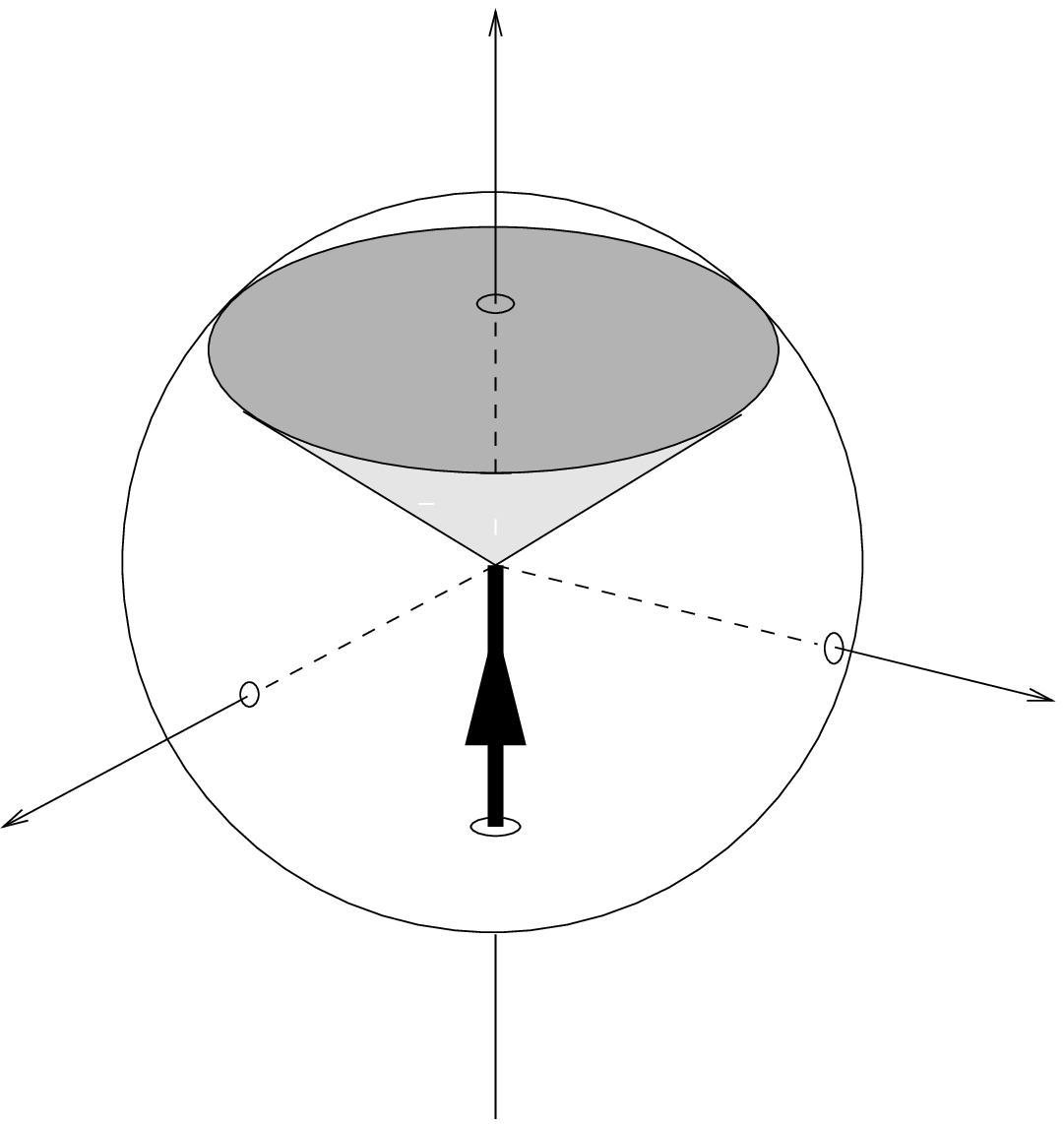, width = 5cm} \\
(c) 
\end{center}
\end{minipage}
\caption{\label{fig4} Illustration of the Dirac type of singularities of
the gauge functions
$V(x)$ which diagonalize the hedgehog type of mapping $\Omega$ (\ref{<36>}) for
various numbers $p$: (a) $p =$ 1, which is the usual hedgehog, where
$V(x)$
has a string singularity at $\vartheta = \pi$, which is the familiar Dirac
string.
(b) $p= 2$, where $V(x)$ is singular in the equatorial plane $\vartheta =
\frac{\pi}{2}$. However, in this plane the (Abelian) magnetic field vanishes and
there is no magnetic monopole in the center of the plane. (c) $p
= 3$, in this case $V(x)$ is singular on the familiar string at $\vartheta =
\pi$
and on the cone defined by $\vartheta = \frac{\pi}{3}$.}
\end{figure}

Fig.\ \ref{fig4} shows the distribution of the (string) singularities of $V (x)$ for
various values of $p$.
The case $p = 1$ is the familiar hedgehog, which has been treated above. In the
case of $p = 2$ the Dirac string is spread out over the equatorial plane. 
However, from the explicit expression of the Abelian magnetic field (\ref{neu1}) one
reads off that the magnetix flux vanishes on this plane. Indeed for $p=2$,
$n=1$, $q=1$ we obtain from (\ref{neu1}) $\bcB(\vartheta=\pi/2)=0$. Thus this
singularity of $V$ does not give rise to a singularity in the magnetic field.
This is because the singular one-dimensional (Dirac) string (in group space) has
been spread out over the whole two-dimensional (equatorial) plane in ordinary
space. 
In the case of $p = 3$ we obtain two types of singularities in ordinary space,
corresponding to $k = 0, 1$.
For
$k = 0$, the area in ordinary space which is mapped onto the singular string in
colour space $\theta = \pi$ is given by a cone defined by 
$\vartheta = \frac{1}{3} \pi$. For $k = 1$, another string singularity
occurs at $\vartheta = \pi$, which is the Dirac string we have already found in 
the field with winding number one. The cone $\vartheta = \pi/3$ can be
considered as a deformation of the equatorial singular plane of the $p = 2$
case. Again this cone carries no magnetic flux.

\end{document}